%% file: conference101719.tex
\documentclass[conference]{IEEEtran}
\IEEEoverridecommandlockouts
 
\usepackage{cite}
\usepackage{amsmath,amssymb,amsfonts}
\usepackage{algorithmic}
\usepackage{pgfplots} % package used to implement the plot  
\usepackage{graphicx}
\usepgfplotslibrary{groupplots}
\usepackage{textcomp}
\usepackage{xcolor}
 \usepackage{tabularx}
\usepackage{subfigure}
\usepackage{caption}
\usepackage{booktabs}
\usepackage{multirow}
\usepackage{listings}
\usepackage{xcolor}

\definecolor{codegreen}{rgb}{0,0.6,0}
\definecolor{codegray}{rgb}{0.5,0.5,0.5}
\definecolor{codepurple}{rgb}{0.58,0,0.82}
\definecolor{backcolour}{rgb}{0.95,0.95,0.92}

\lstdefinestyle{mystyle}{
    backgroundcolor=\color{backcolour},   
    commentstyle=\color{codegreen},
    keywordstyle=\color{magenta},
    numberstyle=\tiny\color{codegray},
    stringstyle=\color{codepurple},
    basicstyle=\ttfamily\footnotesize,
    breakatwhitespace=false,         
    breaklines=true,                 
    captionpos=b,                    
    keepspaces=true,                 
    numbers=left,                    
    numbersep=5pt,                  
    showspaces=false,                
    showstringspaces=false,
    showtabs=false,                  
    tabsize=2,
    frame=single,
    tabsize=2,
    rulecolor=\color{black!30},
    escapeinside={\%*}{*)},
    breaklines=true,
    breakatwhitespace=true,
    framextopmargin=2pt,
    framexbottommargin=2pt,
    extendedchars=false,
    inputencoding=utf8
}

\usepackage{listings}   
\usepackage{caption}
\DeclareCaptionFont{white}{\color{white}}
\DeclareCaptionFormat{listing}{%
  \parbox{\textwidth}{\colorbox{gray}{\parbox{\textwidth}{#1#2#3}}\vskip3pt}}
\lstnewenvironment{tabularlstlisting}[1][]
 {%
  \lstset{}%
 }
 {}

\begin{document}

\title{QML-IDS: Quantum Machine Learning Intrusion Detection System }
%Double Blind
\author{\IEEEauthorblockN{1\textsuperscript{st} Diego Abreu}
\IEEEauthorblockA{
\textit{Federal University of Pará~(UFPA)}\\}
\and
\IEEEauthorblockN{2\textsuperscript{nd} Christian Esteve Rothenberg}
\IEEEauthorblockA{
\textit{University of Campinas~(UNICAMP)}\\}
\and
\IEEEauthorblockN{3\textsuperscript{nd} Antonio Abelém}
\IEEEauthorblockA{
\textit{Federal University of Pará~(UFPA)}\\}
}

%\section*{Agradecimentos}
%Agradecemos ao apoio da Fundação de Amparo à Pesquisa do Estado de São Paulo~(FAPESP), por meio do processo no 2020/04031-1.
\maketitle

\begin{abstract}
The emergence of quantum computing and related technologies presents opportunities for enhancing network security. The transition towards quantum computational power paves the way for creating strategies to mitigate the constantly advancing threats to network integrity. In response to this technological advancement, our research presents QML-IDS, a novel Intrusion Detection System~(IDS) that combines quantum and classical computing techniques. QML-IDS employs Quantum Machine Learning~(QML) methodologies to analyze network patterns and detect attack activities. 
Through extensive experimental tests on publicly available datasets, we show that QML-IDS is effective at attack detection and performs well in binary and multiclass classification tasks. Our findings reveal that QML-IDS outperforms classical Machine Learning methods, demonstrating the promise of quantum-enhanced cybersecurity solutions for the age of quantum utility.

% Through extensive testing on publicly available datasets, we establish QML-IDS's capability in accurately identifying security breaches, showcasing its effectiveness in both binary and multiclass classification scenarios. Our results indicate that qIDS holds its ground against traditional Machine Learning approaches, underscoring the promising role of quantum-driven cybersecurity measures in the burgeoning era of quantum technology.

% The rise of quantum utility in the realm of quantum computing presents not just challenges but also significant opportunities for enhancing network security. This paradigm shift in computational capabilities allows for the development of advanced solutions to counteract the rapidly evolving nature of network attacks. Capitalizing on this technological advancement, this work introduces qIDS, an Intrusion Detection System~(IDS) that innovatively integrates quantum and classical computing approaches. qIDS leverages Quantum Machine Learning~(QML) techniques to effectively learn network behaviors and identify malicious activities. By conducting comprehensive experimental evaluations on public datasets, we demonstrate the proficiency of qIDS in attack detection, excelling in both binary and multiclass classification tasks. Our findings reveal that qIDS competes favorably with classical Machine Learning methods, highlighting the potential of quantum-enhanced cybersecurity solutions in the era of quantum utility.
\end{abstract}


\input{sec/intro}
\input{sec/fundamentacao}
\input{sec/trabrelacionados}

\input{sec/solucaoproposta}
\input{sec/metodologia}
\input{sec/resultados}

\input{sec/conclusao}

% \section*{ACKNOWLEDGMENT}
% The authors would like to thank the São Paulo Research Foundation~(FAPESP) for the financial support through Grant 2020/04031-1.

\bibliographystyle{IEEEtran}
\bibliography{references}

\end{document}

%% file: sec/intro.tex
\section{Introduction}  

Recent advances in quantum computing signify a main shift towards the era of quantum utility, highlighting a crucial phase in the evolution of quantum technologies where quantum computers now outperform classical methods in solving complex problems efficiently and accurately \cite{kim2023evidence}. Despite not yet achieving quantum supremacy—the milestone where quantum computing overtakes classical computing in solving certain tasks within practical timeframes—these developments are crucial strides toward enabling robust applications across various research domains. Within this framework, quantum technologies are poised to revolutionize network security by integrating with Intrusion Detection Systems~(IDS) through Quantum Machine Learning~(QML) techniques. This integration promises more precise detection of network anomalies and the capability to analyze vast datasets, addressing the dynamic challenges of cybersecurity and marking a significant advancement in safeguarding digital infrastructures in the rapidly evolving technological landscape.

% Recent advances in quantum computing research and development mark the entry into the era of quantum utility~\cite{kim2023evidence}, a significant stage in the evolution of quantum technologies. At this stage, quantum computers already demonstrate the ability to perform reliable and efficient calculations, surpassing the limits of classical brute-force methods and providing exact solutions to complex problems. Although quantum supremacy~\cite{boixo2018characterizing}, the point at which a quantum computer can solve problems unreachable for a classical computer in a reasonable time, has not yet been achieved, this current progress represents a fundamental step towards robust practical applications in research, paving the way for a promising future in the use of quantum technologies.

% In this context, quantum technologies offer the possibility of enhancing network security by integrating quantum computing with Intrusion Detection Systems~(IDS) and Intrusion Prevention Systems~(IPS). This advancement can be achieved through the use of Quantum Machine Learning~(QML) techniques~\cite{cerezo2022challenges}, allowing for more accurate detection of suspicious activities on the network, as well as the analysis of large volumes of data, becoming essential in an ever-evolving cybersecurity environment.

A significant challenge is related to the current capabilities of quantum devices, known as Noisy Intermediate-Scale Quantum~(NISQ) devices~\cite{de2022survey}. These limitations include restrictions on the number of qubits~(quantum bits) available, the complexity of the quantum circuits that can be implemented~(depth and available quantum logic gates), and the ability to maintain quantum coherence over time~(due to noise and the inherent nature of qubits)~\cite{gyongyosi2019survey}. Moreover, the lack of robust error correction mechanisms in NISQ systems also represents a significant obstacle~\cite{torlai2020machine}. Therefore, any proposal for a quantum IDS must take these factors into account to ensure its effective applicability in the current quantum computing landscape~\cite{de2022survey}.

In this paper, we propose QML-IDS, a Quantum Machine Learning-Based Attack Detection System. The primary goal is to create an adaptable system for use with NISQ devices, overcoming the inherent limitations of current quantum computing. To achieve this goal, our approach is based on QML techniques, which leverage the capabilities of both classical and quantum computing simultaneously. To evaluate the performance of QML-IDS, a series of experiments were conducted using publicly available network security datasets. Three QML methods in our system were compared in terms of attack detection~(binary classification) and identification of specific attacks~(multiclass classification). Furthermore, the results obtained with QML approaches were compared with classical Machine Learning~(ML) methods. The experimental results provide empirical evidence of the effectiveness of QML techniques in enhancing network attack detection capabilities and point to the feasibility of implementation in NISQ systems. The main contributions of this work are:
\begin{enumerate}    
    \item Development of QML-IDS, a QML-based network attack detection system that utilizes both quantum and classical computing.
    \item Presentation of the operation of QML-IDS through the application of three distinct QML techniques, followed by an evaluation of the performance of each approach.
    \item Implementation of QML-IDS in NISQ systems and evaluation of different quantum circuit configurations on system performance. 
\end{enumerate}

%% file: sec/fundamentacao.tex
\section{Quantum Machine Learning}
Quantum Machine Learning can be understood as a set of techniques that combine principles of quantum computing~(such as superposition, interference, and entanglement) with Machine Learning techniques to perform tasks like classification, regression, and clustering of data. 
\begin{figure*}[h]
    \centering
    \subfigure[Feature Map Circuit.]{
        \label{fig:consumption-rf-oracle}
        \includegraphics[width=0.20\textwidth, height=2.8cm]{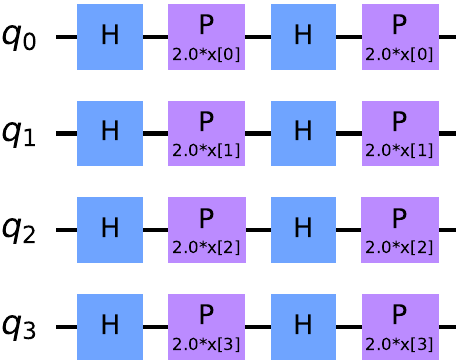}
    }
    \subfigure[Variational Circuit~(Ansatz).]{
        \label{fig:consumption-knn-oracle}
        \includegraphics[width=0.32\textwidth, height=2.8cm]{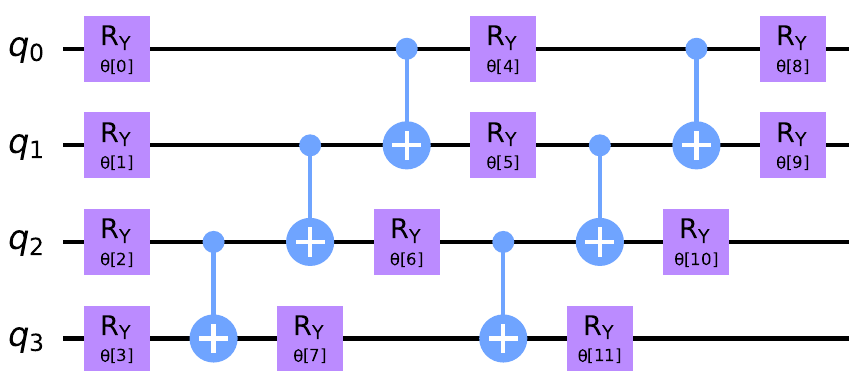}
    }
  \subfigure[QCNN.]{
        \label{fig:consumption-knn-oracle}
        \includegraphics[width=0.42\textwidth, height=3cm]{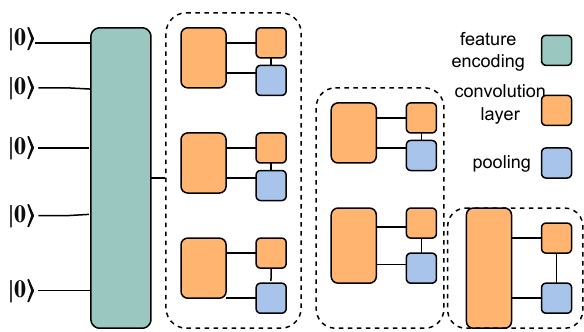}
    }
    \caption{Examples of Quantum Circuits used in QML.}
    \label{fig:cpu-consumption-oracle}
\end{figure*}

\subsection{Variational Quantum Classifier~(VQC)}
The Variational Quantum Classifier~(VQC) \cite{havlivcek2019supervised}  is a hybrid Quantum Machine Learning algorithm that leverages parameterized quantum circuits combined with classical optimization techniques for classification tasks.  It operates by encoding classical input data into quantum states through a process known as a feature map~(Fig 1.a), transforming these states within a quantum circuit whose parameters are iteratively adjusted based on training data to delineate the optimal decision boundary between classes. This encoding and manipulation of quantum states allow the VQC to exploit the quantum mechanical properties of superposition and entanglement, aiming to achieve superior classification performance over classical algorithms in certain scenarios. Thus, the specific choice of the feature map influences the VQC's ability to represent the important features of the input data, directly impacting the performance of the generated model.

The core of the VQC involves three main steps: encoding the input data into quantum states, processing these states through a variational quantum circuit~(or Ansatz, Fig 1.b), and optimizing the circuit parameters using classical optimization techniques. Initially, input data are mapped onto a high-dimensional quantum space using a feature map, which is a unitary operation that prepares the quantum states representing the data. The variational circuit then applies a series of quantum gates, controlled by adjustable parameters, to these states, effectively learning the underlying data patterns.

The final stage involves refining the model through a classical optimization process, where a classical optimizer fine-tunes the \textit{Ansatz} parameters to minimize a cost function. This iterative adjustment aims to identify the optimal circuit configuration that best delineates the target classes. Through this process, the VQC seeks to harness the computational advantages of quantum computing, potentially offering new capabilities in the field of machine learning and data classification.

\begin{figure*}[h]
    \centering
    \includegraphics[width=0.75\textwidth]{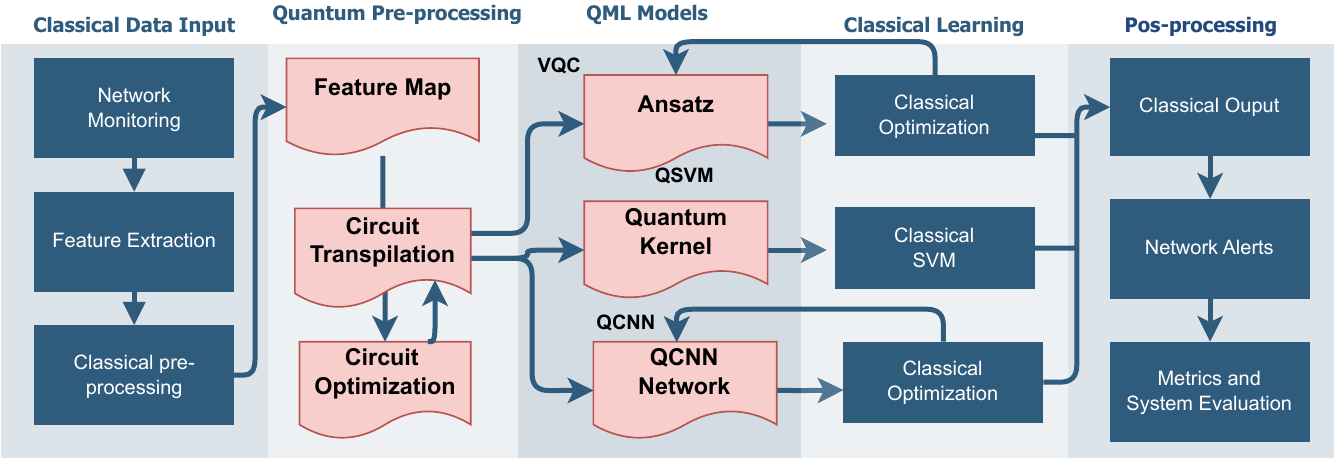}
    \caption{Operational flowchart of QML-IDS.}
    \label{fig:flowchart_qIDS}
\end{figure*}

\subsection{Quantum Support Vector Machines~(QSVM)}

The Quantum Kernel Support Vector Machine~(QSVM) \cite{jager2023universal} represents a quantum-enhanced version of the classical Support Vector Machine~(SVM) algorithm, leveraging the principles of quantum computing to map input data into a high-dimensional quantum feature space. This quantum feature space potentially allows for more effective separation of data that are not linearly separable in their original space. The process involves the construction of a quantum kernel, which is a measure of similarity between pairs of data points in this quantum space. This kernel is generated through quantum transformations applied to the quantum states representing the input data, enabling the QSVM to exploit the computational advantages of quantum mechanics for complex classification tasks.

During the training phase, QSVM utilizes a quantum circuit to project the training examples into the quantum feature space, where it computes the kernel matrix that captures the intricate relationships between these examples. The algorithm then identifies support vectors, which are key data points that define the optimal separation hyperplane in the quantum feature space. These support vectors and the kernel matrix guide the construction of a hyperplane that maximizes the margin between different classes, mirroring the objective of classical SVM but within a quantum computational framework.

In the testing phase, new, unlabeled examples are similarly mapped into the quantum space, and their classification is determined based on their position relative to the quantum hyperplane. This approach allows QSVM to classify data by effectively utilizing quantum operations to handle datasets that challenge traditional classification methods. The effectiveness of QSVM heavily relies on the choice of the feature map and the design of the quantum kernel, which are critical for capturing the essential characteristics of the data in the quantum feature space, thereby enabling the algorithm to achieve high classification accuracy.

 \subsection{Quantum Convolutional Neural Network~(QCNN) }
The Quantum Convolutional Neural Network~(QCNN) \cite{cong2019quantum} is an adaptation of classical convolutional neural networks~(CNNs). The architecture of QCNNs~(Fig 1.c) mirrors that of their classical counterparts, consisting of convolutional layers, pooling layers, and fully connected layers, albeit implemented with quantum operations. The convolutional layers in a QCNN are realized through the application of parameterized unitary operations on neighboring pairs of qubits. These operations are akin to the filters applied in classical CNNs, designed to detect specific features within the data. However, unlike classical filters that operate on pixel values, quantum convolutions manipulate the quantum states of qubits, enabling the extraction of quantum features.

Following the convolutional layers, QCNNs implement quantum pooling layers. The objective of pooling in classical CNNs—to reduce the dimensionality of the data and retain only the most relevant features—is achieved in QCNNs through the measurement of a subset of qubits. This measurement effectively reduces the number of qubits in the system, analogous to the downsampling performed in classical pooling layers. The choice of which qubits to measure and discard is a critical aspect of the QCNN architecture, as it determines how information is condensed and propagated through the network.
Training a QCNN involves adjusting the parameters of the convolutional and fully connected layers to minimize a loss function, similar to the process in classical neural networks. However, due to the quantum nature of the operations, gradient descent and other optimization techniques must be adapted for quantum circuits. This often involves the use of classical optimizers to adjust the quantum parameters based on the measurement outcomes, effectively creating a quantum-classical learning algorithm.
 

%% file: sec/trabrelacionados.tex
\section{Related Work}
%\cite{cerezo2022challenges}
Several studies have addressed the use of Quantum Machine Learning. In this section, we highlight works that seek to utilize QML in the context of network security. Said et al.~(2023) \cite{said2023quantum} explore the application of a QSVM model to detect Distributed Denial of Service~(DDoS) attacks in smart micro-grids. The study evaluates the QML model using a reduced version~(by sampling) of the CIC-DDoS2019 dataset \cite{sharafaldin2019developing}, which includes data on both DDoS attacks and normal network behavior. The results demonstrate the superiority of the QML model compared to the classical approach, SVM. In the paper by Gong et al. \cite{gong2022network}, the application of a Neural Network~(NN) in conjunction with the VQC for network attack detection is proposed. The authors apply the created model to a reduced subset of features from the KDD CUP99 dataset \cite{elkan2000results}, incorporating a class balance approach to increase classification accuracy. Similarly,  Kalinin and Krundyshev \cite{kalinin2023security} propose a QSVM model with NN for network attack detection. The generated model is applied to a database created by the authors and compared with other techniques such as SVM and CNN. These works implemented their solutions using noiseless quantum computing simulators, which cannot accurately represent current NISQ systems. In our work, our proposal is implemented using noisy backends, which more closely represent current NISQ equipment.
%In \cite{suryotrisongko2022adversarial}, the Hybrid Quantum DL is also evaluated in regard to adversarial attack robustness.

Moreover, Suryotrisongko and Musahi \cite{suryotrisongko2022hybrid} propose a VQC and a Hybrid Quantum Deep Learning (DL) to detect botnet domain generation algorithm~(DGA) attacks. The Hybrid DL consists of adding one quantum layer to a NN model. While the DL model was tested using NISQ devices, the VQC was not. 
In our previous research \cite{abreu2024}, QML models were initially tested within a network security context, through direct implementation of baseline circuits. In this study, we expand on that research by employing quantum circuit optimization techniques alongside more thorough experimentation  Thus, this research contributions presented in this work differ by presenting a comprehensive analysis of the application of various QML techniques~(VQC, QSVM, and QCNN), both for detection~(binary scenario) and for the identification of attacks~(multiclass scenario). Additionally, our work conducts tests on three reference datasets in network security, containing a wide variety of network attacks. Furthermore, unlike the related studies, we conducted tests in circuits with diverse optimization levels, incorporating resilience and dynamic decoupling, which contributed to the improvement of our results.

%% file: sec/solucaoproposta.tex
\section{QML-IDS: Hybrid Quantum Machine Learning-Based Attack Detection System}

In this work, we propose a network attack detection system based on Hybrid Quantum Machine Learning~(QML-IDS). The proposal is to apply Hybrid QML techniques, which use both quantum and classical computing, to perform network attack detection. The operation of the proposal is presented in the flowchart in Figure 2.

QML-IDS begins with network monitoring and data collection, from which features are extracted. After this, data preprocessing is performed for normalization, handling missing values, and other techniques aimed at preparing the data for the system. Next, the mapping of classical data to quantum states occurs through the feature map. The Hybrid QML process is then initiated, with the quantum and classical parts according to the QML technique. In the case of VQC, a variational quantum circuit ansatz is generated, whose parameters are adjusted by a classical optimizer. For QSVM, a quantum kernel is generated, which is used to train the classical prediction model. In the QCNN, the quantum network is generated and optimized classically.

%In QKM, the quantum process uses the swap-test circuit to calculate and generate clusters, and a classical optimizer adjusts the parameters of the KQM.

The final result of the process is a classical output, an interpretation of the quantum results that is used for the generation of attack alerts. Based on the analyses performed, the system generates attack alerts whenever suspicious patterns are identified, contributing to network security. It is important to note that in this approach, other hybrid QML methods can also be used, respecting the particularities of each method. QML-IDS can then be implemented partly on a classical system and partly on a quantum system. 

%Due to the absence of commercial quantum computers, the quantum process can be carried out through remote backends, such as quantum simulators available on supercomputers or private quantum computers, like those available from IBM.

% \footnote{https://atos.net/en/lp/myqlm}
% \footnote{https://www.ibm.com/quantum}

%% file: sec/metodologia.tex
\section{Experimental Setup}
To evaluate the proposed system, we conducted a comprehensive case study, leveraging three prominent network security databases: UNSW-NB15~\cite{UNSW-NB15}, CICIDS17~\cite{CIDIDS17}, and CICIoT2023~\cite{neto2023ciciot2023}. These datasets include both normal and various types of attack network traffic data and are frequently utilized for the detection and classification of attacks using ML techniques. Three classical ML methods were selected and will serve as a comparison to the quantum counterparts results: Support Vector Machine~(SVM), Convolution Neural Network~(CNN), and Random Forest~(RF) as a representative classifier to compare with VQC,  all implemented using the scikit-learn and TensorFlow frameworks.

To evaluate the performance of the QML-IDS system, three Quantum Machine Learning models were employed: VQC, QSVM, and QCNN, which were implemented using the Qiskit framework. For these models, the specific hyperparameters used are presented in Table I. We employ four feature maps from Qiskit (RawFeatureVector, PauliFeatureMap, ZFeatureMap, ZZFeatureMap) for encoding classical data into quantum states, alongside four Ansatz and classical optimizers within the Qiskit framework. The RawFeatureVector offers a direct mapping of classical data to quantum states, establishing a baseline. The PauliFeatureMap and ZFeatureMap utilize quantum gates to simulate Pauli operators and employ Hadamard and unitary gates for transforming classical inputs into quantum states, enabling the capture of nonlinear patterns and facilitating first-order data encoding. The ZZFeatureMap extends these capabilities by incorporating second-order qubit interactions for more complex correlation encoding\cite{buonaiuto2023best}.

\begin{table}[h]
\centering
\small
\caption{Hyperparameters of Quantum Models.}
\label{tab:hparams_qml}
\begin{tabular}{lcc}
\toprule
\textbf{Feature Maps}   & \textbf{Ansatz} & \textbf{Optimizer}  \\
\midrule
PauliFeatureMap   & EfficientSU2 & COBYLA   \\
RawFeatureVector  & ExcitationPreserving & ADAM  \\
ZFeatureMap       & RealAmplitudes & SPSA   \\
ZZFeatureMap      & TwoLocal & GradientDescent   \\
\toprule
\textbf{Optimization Level}   & \textbf{Resilience} & \textbf{Decoupling}  \\
\midrule
Level 0,1,2,3   & Level 0 or 1& 0 or 1   \\
\end{tabular}
\end{table}
% \footnote{https://scikit-learn.org/stable/}
% \footnote{https://www.ibm.com/quantum/qiskit}

Addressing the challenge of quantum circuit transpilation for real quantum hardware, our experiments leverage optimization levels from 0 to 3 for circuit adaptation, with each level introducing increasingly sophisticated optimization strategies, from basic gate collapsing to advanced techniques like peep-hole optimization and noise-adaptive qubit mapping. Additionally, we integrate resilience levels and dynamic decoupling to enhance error resilience, with resilience level 0 offering no mitigation and level 1 targeting readout errors through Matrix-free Measurement Mitigation, alongside dynamic decoupling to reduce environmental interactions, thereby optimizing the balance between accuracy and processing time in quantum computations.

%% file: sec/resultados.tex
\section{Results}
\subsection{NISQ \textit{Backend} Results}
In this section, we delve into the performance of QML-IDS across a variety of NISQ \textit{Backend} configurations. For this, six different \textit{backends} were used, including the noise-free quantum computing simulator \textit{QASM} and five real quantum computing environments: \textit{IBM\_CAIRO}, \textit{IBM\_KYOTO}, \textit{IBM\_BRISBANE}, \textit{IBM\_OSAKA}, and \textit{IBM\_SHERBROOKE}. These environments incorporate the specific noise models, quantum logic gates, and topologies of actual quantum computers, providing a realistic assessment of quantum technologies in security applications.

\begin{table}[h]
    \small
    \caption{F1 Score of QML on each \textit{Backend} across three databases~(for binary case) - Part 1.}
    \begin{tabular}{lc|ccc}
    \toprule
    UNSW-NB15 & QASM & IBM\_CAIRO & IBM\_KYOTO \\
    \midrule
    VQC   &88.56\% & 87.91\% & \textbf{88.10}\% \\
    QSVM  &89.34\% & \textbf{87.90}\% & 86.41\% \\
    QNCC  & 87.45\% & 86.12\% & 87.12\% \\
    \toprule
    CIC-IDS-17 & QASM & IBM\_CAIRO & IBM\_KYOTO \\
    \midrule
    VQC   & 95.12\% & 93.60\% & \textbf{94.78}\% \\
    QSVM  & 94.67\% & 92.92\% & \textbf{94.40}\% \\
    QNCC   & 95.88\% & 93.80\% & \textbf{95.60}\% \\
    \toprule
    CICIoT2023 & QASM & IBM\_CAIRO & IBM\_KYOTO \\
    \midrule
    VQC & 82.55\% & 78.55\% & 76.15\% \\
    QSVM & 84.22\% & 79.19\% & 80.00\% \\
    QNCC & 87.81\%  & 82.55\% & \textbf{82.56}\% \\
    \end{tabular}
\end{table}

\begin{table}[h]
    
    \caption{F1 Score of QML on each \textit{Backend} across three databases~(for binary case) - Part 2.}
    \begin{tabular}{lc|ccc}
    \toprule
    UNSW-NB15 & IBM\_BRISBANE & IBM\_OSAKA & SHERBROOKE \\
    \midrule
    VQC   & 85.78\% & 87.75\% & 87.31\% \\
    QSVM  & 86.39\% & \textbf{87.90}\% & 86.55\% \\
    QNCC  & \textbf{87.32}\% & 87.19\% & 85.42\% \\
    \toprule
    CIC-IDS-17 & IBM\_BRISBANE & IBM\_OSAKA & SHERBROOKE \\
    \midrule
    VQC   & 92.40\% & 94.12\% & 94.00\% \\
    QSVM  & 92.80\% & 92.15\% & 94.38\% \\
    QNCC   & 93.40\% & 94.15\% & 94.12\% \\
    \toprule
    CICIoT2023 & IBM\_BRISBANE & IBM\_OSAKA & SHERBROOKE \\
    \midrule
    VQC & 77.12\% & 77.02\% & \textbf{78.92}\% \\
    QSVM & \textbf{82.40}\% &  80.55\% & 80.12\% \\
    QNCC & 82.00\% & 82.11\%  & 81.93\% \\
    \end{tabular}
\end{table}
To illustrate the obtained results, the performance of three QML models~(VQC, QSVM, QNCC) is presented in Tables II and III, showcasing their F1 Scores across the mentioned \textit{backends} for binary attack detection. The F1 Score is chosen for its balanced measure of precision and recall, crucial in security contexts for effectively detecting attacks while minimizing false positives. The \textit{QASM} simulator, representing an ideal quantum computing scenario, consistently shows high F1 Scores across all databases, highlighting the potential of quantum computing in enhancing IDS capabilities. However, the focus of this research is on the performance within real quantum systems, where noise and operational limitations present substantial challenges.

% \begin{table*}[]
%     \caption{F1 Score of QML on each \textit{Backend} across three databases~(for binary case).}
%     \begin{tabular}{lc|ccccc}
%     \toprule
%     UNSW-NB15 & QASM & IBM\_CAIRO & IBM\_KYOTO & IBM\_BRISBANE & IBM\_OSAKA & IBM\_SHERBROOKE \\
%     \midrule
%     VQC   &88.56\% & 87.91\% & \textbf{88.10} \% & 85.78\% & 87.75\% & 87.31\% \\
%     QSVM  &89.34\% & \textbf{87.90}\% & 86.41\% & 86.39\% & \textbf{87.90}\% & 86.55\% \\
%     QNCC  & 87.45\% & 86.12\% & 87.12\% & \textbf{87.32}\% & 87.19\% & 85.42\%\\
%     \toprule
%     CIC-IDS-17 & QASM & IBM\_CAIRO & IBM\_KYOTO & IBM\_BRISBANE & IBM\_OSAKA & IBM\_SHERBROOKE \\
%     \midrule
%     VQC   & 95.12\% & 93.60\% & \textbf{94.78}\% & 92.40\% & 94.12\% & 94.00\%\\
%     QSVM  & 94.67\% & 92.92\% & \textbf{94.40}\% & 92.80\% & 92.15\% & 94.38\%\\
%     QNCC   & 95.88\% & 93.80\% & \textbf{95.60}\% & 93.40\% & 94.15\% & 94.12\% \\
%     \toprule
%     CICIoT2023 & QASM & IBM\_CAIRO & IBM\_KYOTO & IBM\_BRISBANE & IBM\_OSAKA & IBM\_SHERBROOKE \\
%     \midrule
%     VQC & 82.55\% & 78.55\% & 76.15 \% & 77.12\% & 77.02\% & \textbf{78.92}\%\\
%     QSVM & 84.22\% & 79.19\% & 80.00\% & \textbf{82.40}\% &  80.55\% & 80.12\% \\
%     QNCC & 87.81\%  & 82.55\% & \textbf{82.56}\% & 82.00\% & 82.11\%  & 81.93\%\\
%     \end{tabular}
% \end{table*}

Among the real quantum systems, there is a significant variation in performance, which reflects the impact of each system's unique noise model, quantum logic gates, and topology. This variation underscores the importance of selecting and tuning QML models according to the specific characteristics of the quantum hardware in use. For instance, VQC demonstrates robust performance on the \textit{IBM\_KYOTO} backend for the CICIDS17 dataset, while QSVM shows adaptability with strong results on \textit{IBM\_OSAKA} for the UNSW-NB15 database and on \textit{IBM\_BRISBANE} for the CICIoT2023 database. QNCC, in particular, stands out with the highest F1 Scores in several configurations, such as on the \textit{IBM\_KYOTO} and \textit{IBM\_BRISBANE} backends for the CICIDS17 and CICIoT2023 databases, respectively, showcasing its effectiveness in handling the complexities of real quantum systems. 

The results indicate the potential of QML-IDS to leverage quantum computing for security applications, while also highlighting the critical role of hardware-specific considerations in optimizing performance. The varying results across different \textit{backends} and models emphasize the necessity for ongoing research and development in quantum computing to address the challenges posed by noise and other physical limitations in NISQ systems. The best results obtained in each model and \textit{backend} combination will be discussed in detail in the following subsections.

\subsection{QML-IDS Results: Attack Detection}
Table IV provides a detailed comparison of Quantum Machine Learning models against traditional Machine Learning, in the context of binary classification for attack detection. The comparison spans three distinct datasets: UNSW-NB15, CIC-IDS-17, and CICIoT2023, showcasing the F1 score as a metric to evaluate the performance of each method. Notably, QML models, including Variational Quantum Classifier, Quantum Support Vector Machine, and Quantum Convolutional Neural Network, demonstrate competitive or superior performance compared to their ML counterparts across all datasets.

\begin{table}[h]
    \caption{F1 score results for the binary classification.}
    \label{tab:performance_comparison_2}
    \begin{tabular}{l|c|c|c}
   
   QML & UNSW-NB15 & CIC-IDS-17 & CICIoT2023 \\
    \midrule
    VQC   & 88.10\% & 94.78\%& 78.92\% \\
    QSVM  &  87.90\%& 94.40\% & 82.40\% \\
    QCNN   & 87.32\% & 95.60\% & 82.56\%\\
    \midrule
    RF    & 82.67\% & 92.45\% & 71.95\% \\
    SVM   & 82.34\% & 93.78\% & 82.7\%  \\
    CNN   & 86.72\% & 93.15\% & 78.52\%\\
    \end{tabular}
\end{table}

The performance of VQC is particularly impressive, marking the highest F1 scores among QML models across the datasets, which underscores its effectiveness in detecting attacks. This is evident in the comparison where VQC achieves an F1 score of 88.10\% on the UNSW-NB15 dataset, surpassing the best performing traditional ML method, SVM, which scores 82.34\%. Similar trends are observed in the CIC-IDS-17 and CICIoT2023 datasets, where QML models generally outperform traditional ML methods, albeit with varying margins. QSVM and QCNN also show strong performance, with QCNN achieving the highest F1 score of 95.60\% on the CIC-IDS-17 dataset, indicating the potential of QML models in enhancing cybersecurity measures.

% Observing the results, it's notable that QMLs, in general, show comparable or superior performance to ML methods. In particular, VQC stands out with the highest F1 Scores across all three datasets, indicating significant effectiveness in attack detection. For example, in the UNSW-NB15 dataset, VQC achieves 88.99\%, while the best ML method, KNN, reaches 84.89\%. This trend is consistent in CIC-IDS-17 and TON-IOT, where QMLs outperform ML methods, although by varying margins.

% An important observation is that, while QMLs show advantages in some cases, the performance difference between them and traditional ML methods is not significantly large. This suggests that, while QMLs offer potential benefits, especially in specific scenarios or when properly adjusted to the context and data characteristics, they are not yet substantially superior in all aspects.

\subsection{qIDS Results: Identification of Attacks}
\begin{table}[h]
    \small
    \caption{Multiclass F1 score Results: UNSW-NB15 dataset.}
    \label{tab:attack_comparison}
    \begin{tabular}{lccccccc}
   
    \textbf{Attack} & \textbf{VQC} & \textbf{QSVM} & \textbf{QCNN} & \textbf{RF} & \textbf{SVM} & \textbf{CNN} \\
    \midrule
    Analysis   & 95.55 & 87.85 & \textbf{98.88}  & 90.23 & 96.45 & 95.15 \\
    Backdoor   & 84.16 & 82.08 & 91.05  & \textbf{92.11} & 90.78 & 90.13 \\
    DoS        & 89.29 & 88.68 & \textbf{93.23} & 82.56  & 92.45 & 80.26 \\
    Exploits   & 95.23 & 93.78 & \textbf{94.44}  & 85.89  & 79.01& 78.16 \\
    Fuzzers    & 76.12 & \textbf{96.55} & 93.14  & 79.78  & 77.56& 76.89 \\
    Generic    & 97.01 & 94.45 & 95.00  & 99.01  & \textbf{99.34}& 94.83 \\
    Recon     & 98.88 & \textbf{99.72} & 97.48  & 82.89  & 86.12& 90.46\\
    Shellcode  & 68.23 & 76.67 & 79.94  & \textbf{80.12}  & 76.23 & 72.00 \\
    Worms      & 22.86 & 55.89 & \textbf{65.11}  & 35.78 & 20.56 & 48.15
    \end{tabular}
\end{table}

\begin{table}[h]
    \caption{Multiclass F1 score Results: CIC-IDS-17 dataset.}
    \label{tab:desempenho_modelos}
    \begin{tabular}{lccccccc}

    \textbf{Attack} & \textbf{VQC} & \textbf{QSVM} & \textbf{QCNN} & \textbf{RF} & \textbf{SVM} & \textbf{CNN} \\
    \midrule
    BoT            & 88.23   & \textbf{97.11}   &   98.45    & 92.12 & 94.67   & 91.45   \\
    BruteForce     & 95.87   & 99.99  &  \textbf{99.99}    & 96.24   & 93.45  &  94.42  \\
    DoS            & 90.32   & 90.79   & 92.45      & 98.02  & \textbf{99.67}  &  96.15 \\
    DDoS           & 90.54   & \textbf{99.99}  & 96.55     & 99.34 & 94.32  & 96.32 \\
    Infiltration   & 90.99   & \textbf{99.66}   & 97.42     & 97.53  & \textbf{99.87}  & 97.15   \\
    PortScan       & 95.45   & 96.32   &  \textbf{97.48}     & \textbf{97.45}  & \textbf{97.89}  &  96.88  \\
    WebAttack      & 92.78   & 96.67   &  97.83   & 96.89  & \textbf{98.21}   &  96.55  \\
 
    \end{tabular}
\end{table}
\begin{table}[h!]
    \caption{Multiclass F1 score Results: CICIoT2023 dataset. }    \label{tab:ton_iot_attack_comparison}
    \begin{tabular}{lccccccc}
    \textbf{Attack} & \textbf{VQC} & \textbf{QSVM} & \textbf{QCCN} & \textbf{RF} &  \textbf{SVM}  & \textbf{CNN} \\
    \midrule
    BruteForce & 64.19 & 70.25 & \textbf{74.76} & 62.55 & 60.41 & 66.31\\
    DoS   & 92.03 & 95.59 & \textbf{96.43} & 95.13  & 92.86 & 94.41      \\
    DDoS  & 92.81 & 95.48 & \textbf{96.37} & 95.78  & 92.52 & 95.27     \\
    Mirai  & 90.32 & 96.92 & \textbf{97.48} & 96.44 & 93.00  & 97.92  \\
    Recon  & 83.40 & 80.60 & \textbf{92.53}  & 85.88 & 81.95& 88.44        \\
    Spoofing & 74.95 & 70.20 & \textbf{76.08}  & 65.73 & 60.69 & 68.27    \\
    Web & 60.00 & 72.85 & \textbf{74.56}  & 60.48  & 58.10 & 68.63  \\
    \end{tabular}
\end{table}

The multiclass classification results for attack identification across the UNSW-NB15, CIC-IDS-17, and CICIoT2023 datasets, as shown in Tables V, VI, and VII, reveal insightful trends about the capabilities of Quantum Machine Learning~(QML) models versus traditional Machine Learning~(ML) methods. In the UNSW-NB15 dataset, QML models such as VQC, QSVM, and QCNN display varied performance across different attack types. They show exceptional proficiency in identifying attacks like Analysis and Reconnaissance, where QCNN notably excels with scores reaching up to 98.88\% for Analysis attacks. However, for categories like Worms and Shellcode, these quantum models lag behind, suggesting that while QMLs are promising for certain attack vectors, traditional ML methods like SVM and RF still hold the upper hand in others.

Moving to the CIC-IDS-17 dataset, the pattern of QML models, particularly QSVM and QCNN, achieving high F1 Scores in detecting specific types of attacks such as BruteForce and DDoS is evident. This indicates their potential in accurately identifying these attacks, with QSVM and QCNN reaching near-perfect scores in BruteForce detection. Yet, in scenarios involving DoS and WebAttack, traditional methods like SVM and RF present competitive or superior performance, highlighting the nuanced effectiveness of QML models depending on the nature of the attack. The analysis of the CICIoT2023 dataset further underscores the strengths and limitations of QML in the realm of cybersecurity. Here, QML models demonstrate robust performance in identifying DoS attacks, with QCNN showing a remarkable F1 Score of 96.43\%. However, for other attack types such as Spoofing and Web attacks, the performance of QML models is less dominant, with traditional ML methods occasionally outperforming or matching the quantum approaches. This comparative analysis across three datasets illustrates the evolving landscape of intrusion detection, where QML models offer significant advantages for certain attack types but still require advancements to consistently outperform traditional ML methods across the board.

Therefore, despite the promising results, the performance gap between QML and traditional ML methods is not overwhelmingly large, suggesting that while QML offers potential advantages in certain scenarios, it does not yet decisively outperform traditional approaches in all aspects. This highlights the importance of further research and development in QML to fully exploit its capabilities and potentially achieve substantial improvements over traditional ML methods in the field of cybersecurity.

%% file: sec/conclusao.tex
\section{Conclusion and Future Work}

This work introduced QML-IDS, a Hybrid Quantum Machine Learning-based attack detection system, designed to tackle emerging challenges in cybersecurity scenarios. Through experimental evaluations conducted on different public datasets, QML-IDS proved to be effective in detecting attacks, both in binary and multiclass classifications, showing competitive results compared to traditional Machine Learning methods.

In the current NISQ and quantum utility scenarios, the application of qIDS already demonstrates competitive results compared to other ML techniques in the detection and identification of attacks. However, there are still challenges and limitations to be overcome for the practical implementation of the proposal in real network environments, which continues to be an active area of research and opens up space for future works. Among the points to be explored are integration with existing classical IDS frameworks and with traditional ML-based IDS systems to create a more robust defense mechanism. Moreover, addressing the constraints of quantum hardware availability and navigating the privacy concerns associated with processing data on platforms like IBM's will be crucial in pushing the boundaries of quantum-enhanced cybersecurity solutions.

\section*{Acknowledgements}
This work was partially supported by the São Paulo Research Foundation (FAPESP), project 2020/04031-1, project 2023/00811‑0, project 2021/00199-8, CPE SMARTNESS, and project 2018/23097-3. It was also financed in part by Coordination of Superior Level Staff Improvement (CAPES) Finance Code 001